\documentclass[aip,jcp,reprint]{revtex4-1}

\usepackage{graphicx}
\usepackage{amssymb,amsfonts,amsmath}
\usepackage{epsf}
\usepackage{subfigure}
\usepackage{epstopdf}
\usepackage{mathrsfs}
\usepackage{color}

\newcommand{\be}{\begin{equation}}
\newcommand{\ee}{\end{equation}}
\newcommand{\bea}{\begin{eqnarray}}
\newcommand{\eea}{\end{eqnarray}}

\begin{document}

\title{Nonequilibrium thermodynamics and boundary conditions \\ for reaction and transport in heterogeneous media}

\author{Pierre Gaspard}
\email{gaspard@ulb.ac.be}
\affiliation{Center for Nonlinear Phenomena and Complex Systems, Universit{\'e} Libre de Bruxelles, Code Postal 231, Campus Plaine, B-1050 Brussels, Belgium}
\author{Raymond Kapral}
\email{rkapral@chem.utoronto.ca }
\affiliation{Chemical Physics Theory Group, Department of Chemistry, University of Toronto, Toronto, Ontario M5S 3H6, Canada}

\begin{abstract}
Nonequilibrium interfacial thermodynamics is formulated in the presence of surface reactions for the study of diffusiophoresis in isothermal systems.  As a consequence of microreversibility and Onsager-Casimir reciprocal relations, diffusiophoresis, i.e., the coupling of the tangential components of the pressure tensor to the concentration gradients of solute species, has a reciprocal effect where the interfacial currents of solutes is coupled to the slip velocity.  The presence of surface reactions is shown to modify the diffusiophoretic and reciprocal effects at the fluid-solid interface.  The thin-layer approximation is used to describe the solution flowing near a reactive solid interface.  Analytic formulas describing the diffusiophoretic and reciprocal effects are deduced in the thin-layer approximation and tested numerically for the Poiseuille flow of a solution between catalytic planar surfaces.
\end{abstract}

\maketitle

\section{Introduction}

Interfacial phenomena play a crucial role in many nonequilibrium systems and this is especially the case for the motion of colloidal particles through phoretic mechanisms.  In these mechanisms, a force is exerted on particles due to gradients in fields such as concentration, temperature, or electrochemical potential.  This results in a flow in the surrounding fluid, which is responsible for the motion of the colloid.  The interaction of the fluid with the particle through boundary conditions applied at the fluid-solid interface is an essential element in these phoretic mechanisms.\cite{DD74,A86,A89,AP91}

Interest in this area has increased recently with the observation of the self-propulsion of micro-motors through self-phoresis. In this case, the gradient is generated by the particle itself as a result of asymmetric activity on the motor-fluid interface. The situation where the activity arises from chemical reactions on a portion of the motor surface is of special interest.  This asymmetric catalytic activity can lead to gradients giving rise to diffusiophoresis, electrophoresis, and thermophoresis.  For instance, some of the first nanomotors were bimetallic rods that catalyzed the decomposition of hydrogen peroxide to produce gradients in electrochemical potentials that drove the motion.\cite{PKOSACMLC04,FBAMO05}  Motors with other geometries featuring catalytic and noncatalytic parts have been studied extensively.\cite{W13,WDAMS13,SSK15}  Self-powered micropumps based on these mechanisms have also been built.\cite{SPOASDCMS14,DWDYMS15,YDBS15}  In all of these cases, the form that the boundary conditions take is an important ingredient in the continuum description of the mechanism underlying propulsion.\cite{A86,A89,AP91}  A particularly simple geometry is a spherical Janus motor with reactive and inactive hemispheres, which may be propelled by self-diffusiophoresis involving concentration gradients of reactants and products.\cite{GLA07,K13} The diffusiophoretic mechanism is the main focus of this paper.  In a more general context, these issues also concern transport in a condensed phase close to a reactive surface and its effect on the reaction itself.

These considerations have motivated our examination of the boundary conditions at an interface where reaction and transport occurs.  Much effort has already been devoted to the general problem of boundary conditions at an interface.\cite{HB73}  The approach we take is based on nonequilibrium interfacial thermodynamics,\cite{W67,BAM76,K77,B86,KB08} which is developed here to treat surface reactions of multicomponent solutions.  The boundary conditions we derive for the velocity and concentration fields can be applied to a variety of different situations, ranging from self-propulsion of active colloids to the effects of fluid flows on catalytic reactive surfaces.

The entropy balance equation is established from the Gibbs and Euler thermodynamic relations, and it allows us to identify the different interfacial irreversible processes beyond those that are known for bulk phases.\cite{P67,H69,GM84,N79}  Following the principles of linear nonequilibrium thermodynamics, phenomenological relations are formulated between the currents and affinities of the different processes. As a consequence of microreversibility, the linear response coefficients obey Onsager-Casimir reciprocal relations.\cite{O31a,O31b,C45}  It is shown that the balance equations for the surface concentrations of the different species and their coupling to the velocity field provide the route to obtain the boundary conditions. In the case of a fluid-solid interface, the molecular species in the solution are assumed to interact with the solid surface through potential energy functions of finite range. When the range is small with respect to the other characteristic lengths of the problem, the thin-layer approximation can be used to determine the boundary conditions. Through this procedure, we obtain the diffusiophoretic and reaction rate constants, and show that the presence of reaction modifies previously obtained expressions for these constants.\cite{A89,AP91,AB06}

The paper is organized as follows.  Section~\ref{thermo} sets up interface nonequilibrium thermodynamics for surface reactions coupled to hydrodynamic flows in isothermal multicomponent systems, especially at fluid-solid interfaces.  In Sec.~\ref{ThinLayer}, the desired boundary conditions are deduced within the thin-layer approximation and the modified diffusiophoretic and reaction rate constants are obtained.  Numerical examples are given in Sec.~\ref{Ex}.  Section~\ref{Conclusion} contains the conclusions and gives perspectives on the study.

\section{Interfacial nonequilibrium thermodynamics}
\label{thermo}

\subsection{General formulation}

We first present the general formulation of nonequilibrium interfacial thermodynamics. With this aim in mind, we consider an interface between two bulk phases and include irreversible processes due to chemical reactions -- especially at the interface -- within the framework of Refs.~\onlinecite{W67,BAM76,K77,B86,KB08}.  Since we are mainly interested in diffusiophoresis, we include the concentration fields of all solute species in the description.  Furthermore, we suppose that the system is isothermal, so that all terms involving thermal non-uniformities are eliminated.

The possibly moving interface is located at $f({\bf r},t)=0$, and it divides the system in two media, $+$ at $f({\bf r},t)>0$ and $-$ at $f({\bf r},t)<0$.  These media can be fluids or solids, but at least one of them is a fluid.  The vector normal to the interface is denoted ${\bf n}\equiv\pmb{\nabla}f({\bf r},t)/\Vert\pmb{\nabla}f({\bf r},t)\Vert$.  It is convenient to introduce a Dirac delta distribution located at the interface defined by
\be
\delta^{\rm s}({\bf r},t) \equiv \Vert\pmb{\nabla}f({\bf r},t)\Vert \, \delta\left[f({\bf r},t)\right] ,
\label{delta-s}
\ee
as well as Heaviside indicator functions for both bulk phases $\theta^{\pm}({\bf r}) \equiv \theta\left[\pm f({\bf r},t)\right]$.  Any density $x$ can be decomposed as
\be
x = x^+ \theta^+ + x^{\rm s} \delta^{\rm s} + x^- \theta^-,
\label{decomp-x}
\ee
where $x^{\pm}$ are the values of the quantity $x$ in the two bulk phases on both sides of the interface, while $x^{\rm s}$ is the excess surface density of $x$.  As a consequence of these definitions, Refs.~\onlinecite{W67,BAM76,K77,B86,KB08} show that the bulk and surface densities obey balance equations for mass, momentum, energy, and entropy for media composed of a single atomic or molecular species.

Here, we consider molecular mixtures composed of several species $k=1,2,...$. The balance equations for the concentration $c_k$ of species $k$ are given by
\bea
&& \partial_t c_k^{\pm} + \pmb{\nabla}\cdot\left( c_k^{\pm} {\bf v}^{\pm} + {\bf j}_k^{\pm}\right) = \sum_{r}\nu_{kr}w_{r}^{\pm}  , \\
&& \partial_t \Gamma_k  + \pmb{\nabla}_{\bot}\cdot\left( \Gamma_k {\bf v}^{\rm s} + {\bf j}_k^{\rm s}\right) = \sum_{r}\nu_{kr}w_{r}^{\rm s}  - {\bf n}\cdot{\bf j}_k^+ + {\bf n}\cdot{\bf j}_k^-  , \quad\label{bal-eq-Gk}\\
&& {\bf n}\cdot {\bf j}_k^{\rm s}= 0  ,
\eea
where ${\bf v}^{\pm}$ and ${\bf v}^{\rm s}$ are the bulk and surface fluid velocities, ${\bf j}_k^{\pm}$ and ${\bf j}_k^{\rm s}$ the bulk and surface current densities of species~$k$, $w_{r}^{\pm}$ and $w_{r}^{\rm s}$ the bulk and surface reaction rates~$r$, $\nu_{kr}$~the stoichiometric coefficient of species~$k$ in reaction~$r$, $\Gamma_k\equiv c_k^{\rm s}$ the excess surface density of species~$k$, and $\pmb{\nabla}_{\bot}$ is the tangential gradient.  If positive, the variables $\Gamma_k$ characterize the adsorption of species $k$ at the interface.\cite{RW02}  However, $\Gamma_k$ may be negative if there is a deficit of species $k$ at the interface.  The bulk mass density is related to the concentrations and the molecular masses $m_k$ according to $\rho=\sum_km_kc_k$, where the sum includes the solvent.

The entropy balance equation can be established by assuming local equilibrium, implying the validity of the Gibbs and Euler relations in the bulk phases as well as at the interface. In the bulk phases we have
\be
de^{\pm}=T^{\pm}ds^{\pm}+\sum_k\mu_k^{\pm} dc_k^{\pm}  , \  e^{\pm}=T^{\pm}s^{\pm}-P^{\pm}+\sum_k\mu_k^{\pm}c_k^{\pm}  ,
\ee
where $e^{\pm}$ are the internal energy densities, $T^{\pm}$ the temperatures, $s^{\pm}$ the entropy densities, $\mu_k^{\pm}$ the chemical potentials of species $k$, and $P^{\pm}$ the hydrostatic pressures. The analogous equations at the interface are
\be
de^{\rm s}=T^{\rm s}ds^{\rm s}+\sum_k\mu_k^{\rm s} d\Gamma_k  , \quad e^{\rm s}=T^{\rm s}s^{\rm s}+\gamma+\sum_k\mu_k^{\rm s}\Gamma_k  ,
\ee
where $T^{\rm s}$ is the surface temperature, $\mu_k^{\rm s}$ the surface chemical potential of species $k$, and $\gamma=-P^{\rm s}$ the surface tension defined as minus the hydrostatic surface pressure.  The entropy balance equations are thus given by
\bea
&& \partial_t s^{\pm} + \pmb{\nabla}\cdot\left( s^{\pm} {\bf v}^{\pm} + {\bf j}_s^{\pm}\right) = \sigma_s^{\pm} \, , \\
&& \partial_t s^{\rm s} + \pmb{\nabla}\cdot\left( s^{\rm s} {\bf v}^{\rm s} + {\bf j}_s^{\rm s}\right) = \sigma_s^{\rm s} - {\bf n}\cdot{\bf j}_s^+ + {\bf n}\cdot{\bf j}_s^-   \, , \\
&& {\bf n}\cdot {\bf j}_s^{\rm s}= 0 \, ,
\eea
expressed in terms of the bulk entropy current density ${\bf j}_s^{\pm}$ and entropy production per unit time and unit volume $\sigma_s^{\pm}$ given in Ref.~\onlinecite{GM84}, the excess surface entropy current density ${\bf j}_s^{\rm s}$, and the excess surface entropy production per unit time and unit area $\sigma_s^{\rm s}$.  If the system is isothermal with the common temperature $T\equiv T^+=T^-=T^{\rm s}$, the excess surface entropy current density is given by ${\bf j}_s^{\rm s}=-\sum_k \mu_k^{\rm s}\, {\bf j}_k^{\rm s}/T$ and the excess surface entropy production can be expressed as
\be
\sigma_s^{\rm s} = \sum_\alpha A_\alpha \, J_\alpha
\label{entrprod}
\ee
in terms of the affinities and currents given in Table~\ref{table1}.

\begin{table*}
\caption{\label{table1} The irreversible processes at an isothermal interface. $(\pmb{\nabla}{\bf v}^{\rm s})^{\rm sym}$ denotes the symmetrized gradient of the surface velocity, ${\boldsymbol{\mathsf 1}}_{\bot}\equiv {\boldsymbol{\mathsf 1}}-{\bf n}{\bf n}$, $\stackrel{\circ}{\pmb{\Pi}^{\rm s}}$ is the traceless part of the viscous surface pressure tensor, and $\Pi^{\rm s}$ half its trace.\cite{BAM76}
}
\vspace{5mm}
\begin{center}
\begin{tabular}{|lllcc|}
\hline
irreversible processes & affinity  $A_{\alpha}$ & current $J_{\alpha}$ & space & time\\
\hline
 & & & & \\
dilatational interfacial viscosity &  $A_{\rm p}=-\frac{1}{T} \, \pmb{\nabla}\cdot {\bf v}^{\rm s}$ & $J_{\rm p}=\Pi^{\rm s}$ & scalar & odd \\
interfacial reaction $r$ & $A_r=-\frac{1}{T} \sum_{k} \mu_k^{\rm s} \, \nu_{kr}$ & $J_r=w_{r}^{\rm s}$& scalar & even \\
transport of species $k$ across interface & $A_{k\bot}=-\frac{1}{T} \left(\mu_k^+-\mu_k^-\right)  $ & $J_{k\bot}=\frac{1}{2} {\bf n}\cdot\left({\bf j}_k^++{\bf j}_k^-\right)$ & scalar & even \\
transport of species $k$ to interface & $A_{k\top}=-\frac{1}{T} \left(\frac{\mu_k^++\mu_k^-}{2}-\mu_k^{\rm s}\right)$ & $J_{k\top}={\bf n}\cdot\left({\bf j}_k^+-{\bf j}_k^-\right)$ & scalar & even \\
transport of species $k$ inside interface & ${\bf A}_{k\parallel}=-\frac{1}{T} \, \pmb{\nabla}_{\bot} \mu_k^{\rm s}$ & ${\bf J}_{k\parallel}={\bf j}_k^{\rm s}$ & vector & even \\
interfacial slippage & ${\bf A}_{\rm v}=-\frac{1}{T}({\bf v}^+-{\bf v}^-)$  & ${\bf J}_{\rm v}=\frac{1}{2}\, {\bf n}\cdot\left({\boldsymbol{\mathsf P}}^++{\boldsymbol{\mathsf P}}^-\right)$ & vector & odd \\
shear interfacial viscosity & $\stackrel{\circ}{{\boldsymbol{\mathsf A}}}_{\rm p}=-\frac{1}{T}\, {\boldsymbol{\mathsf 1}}_{\bot}\cdot\left[(\pmb{\nabla}{\bf v}^{\rm s})^{\rm sym}-\frac{1}{2}\pmb{\nabla}\cdot{\bf v}^{\rm s}\right]\cdot{\boldsymbol{\mathsf 1}}_{\bot}$ & $\stackrel{\circ}{{\boldsymbol{\mathsf J}}}_{\rm p}=\stackrel{\circ}{\pmb{\Pi}^{\rm s}}$ & tensor & odd \\
 & & & & \\
\hline
\end{tabular}
\end{center}
\end{table*}

We suppose that the irreversible processes are driven in their linear regime close to thermodynamic equilibrium. We apply the Curie symmetry principle by considering an isotropic surface.  Accordingly, the currents and the affinities of the same spatial character are linearly related to each other by
\be
J_{\alpha}= \sum_{\beta} L_{\alpha\beta} A_{\beta } \, .
\ee
The response coefficients obey the Onsager-Casimir reciprocal relations
\be
L_{\alpha\beta} = \epsilon_{\alpha}\epsilon_{\beta} L_{\beta\alpha} \, ,
\label{OCrr}
\ee
where $\epsilon_{\alpha}=\pm 1$ depending on whether the affinity $A_{\alpha}$ is even or odd under time reversal.\cite{BAM76,GM84}  An important result~\cite{GM84} is that the antisymmetric linear response coefficients, $L_{\alpha\beta} = -L_{\beta\alpha}$ relating processes with opposite parities under time reversal, do not appear in the quadratic expression of the entropy production~(\ref{entrprod}).

In specific systems, there is no need to consider all of the linear response coefficients for every possible coupling between the irreversible processes.  In particular, the transport to the interface could be fast enough so that there is a quasiequilibrium between the interface and the bulk phases, in which case the chemical potential of the interface is equal to those of the bulk phases, $\mu_k^{\rm s}=\mu_k^+=\mu_k^-$, and there is no longer transport across and to the interface.  The conditions for this quasiequilibrium is that the linear response coefficients relating $J_{k\bot}$~to~$A_{k\bot}$ and $J_{k\top}$~to~$A_{k\top}$ in Table~\ref{table1} should be large enough to reduce chemical potential differences to a level smaller than the differences between the bulk chemical potentials driving the surface chemical reactions.  If this is not the case, the concentration profiles may manifest a jump across the interface under nonequilibrium conditions, as in the phenomenon of Kapitza thermal resistance;\cite{WTKvEB15} moreover, with the possibility of the interfacial accumulation of adsorbates under equilibrium or nonequilibrium conditions.\cite{RW02}

\subsection{Fluid-solid interface}

Diffusiophoresis is one of the key processes for a fluid-solid interface.  Here, our aim is to determine the boundary conditions associated with this process on the basis of nonequilibrium interfacial thermodynamics and in the presence of interfacial reactions.  While such boundary conditions have been considered previously,\cite{A89,AP91,AB06} our results provide a more general formulation that allows one to describe the effects of reaction on diffusiophoresis.

In fluid-solid systems, the interface can be assumed to be rigid enough so that there is no need to consider the interfacial viscosities.  Similarly, for a solution in contact with a solid, there is no transport across the interface, so that the transport to the interface is only from the solution.  We thus remain with the relations between the scalar reaction rate~$J_r$ and the chemical affinity~$A_r$, and those for the vectorial quantities in Table~\ref{table1}.  According to Ref.~\onlinecite{BAM76}, and using the further boundary condition\cite{BAM77} ${\bf n}\cdot{\boldsymbol{\mathsf P}}^+={\bf n}\cdot{\boldsymbol{\mathsf P}}^-$, the phenomenological surface relations for the vectorial quantities that are consistent with the second law are given by
\bea
{\bf n}\cdot{\boldsymbol{\mathsf P}}\cdot{\boldsymbol{\mathsf 1}}_{\bot} &=& -\frac{L_{\rm vv} }{T}\, {\bf v}_{\rm slip}  - \sum_k \frac{L_{{\rm v}k}}{T}\, \pmb{\nabla}_{\bot}\mu_k^{\rm s}\,  ,  \qquad \label{BC3}\\
{\bf j}_k^{\rm s} &=& -\frac{L_{k{\rm v}}}{T}\, {\bf v}_{\rm slip}  - \sum_l \frac{L_{kl}}{T}\, \pmb{\nabla}_{\bot}\mu_l^{\rm s}\, , \label{BC2}
\eea
where ${\bf v}_{\rm slip}={\bf v}^+-{\bf v}^-$ is the slip velocity between the fluid and the solid, $\mu_k^{\rm s}$ is the surface chemical potential of species $k$, ${\boldsymbol{\mathsf 1}}_{\bot}\equiv{\boldsymbol{\mathsf 1}}-{\bf n}{\bf n}$,  and $\boldsymbol{\mathsf P}$ denotes the fluid pressure tensor at the boundary.  This latter quantity is given for an incompressible fluid by
\be
P_{ij} = P \, \delta_{ij} -\eta (\partial_i v_j + \partial_j v_i) \, ,
\ee
where $P$ is the hydrostatic pressure and $\eta$ the shear viscosity of the fluid phase.  Moreover, the coefficient of sliding friction  $\lambda$ enters through the relation\cite{BB13} $L_{\rm vv} = T  \lambda$, while the coefficients $L_{kl}=L_{lk}$ in Eq.~(\ref{BC2}) are related to the surface diffusion coefficients of the adsorbates if they exist.  For a dilute interfacial coverage of adsorbates, the surface chemical potential of species~$k$ are given by $\mu_k^{\rm s}=\mu_k^{{\rm s}0}+k_{\rm B}T \ln(\Gamma_k/\Gamma^0)$ if $\Gamma_k>0$, so that we recover Fick's law for surface diffusion if $L_{kl}/T = \delta_{kl} D_{k}^{\rm s} \Gamma_k/(k_{\rm B}T)$, where the $D_{k}^{\rm s}$ denote the surface diffusion coefficients.

When the boundary layer and neighboring bulk phase are in local equilibrium the surface chemical potentials take their bulk values: $\mu_k^{\rm s}=\mu_k=\mu_k^0 +k_{\rm B}T \ln(c_k/c^0)$.  The hydrostatic pressure $P$ can be eliminated from Eq.~(\ref{BC3}) because ${\bf n}\cdot{\boldsymbol{\mathsf 1}}_{\bot}=0$.  Under these conditions, and introducing the slip length
\be
b\equiv \frac{\eta}{\lambda} \, ,
\label{b}
\ee
and diffusiophoretic constants~\cite{A89,AP91}
\be
b_k \equiv  k_{\rm B}  \, \frac{L_{{\rm v}k}}{\lambda c_k} \, ,
\label{bk-L}
\ee
Eq.~(\ref{BC3}) takes the simpler form,
\be
b \, {\bf n}\cdot\left(\pmb{\nabla}{\bf v}+\pmb{\nabla}{\bf v}^{\rm T}\right)\cdot{\boldsymbol{\mathsf 1}}_{\bot}={\bf v}_{\rm slip} + \sum_k b_k \, \pmb{\nabla}_{\bot}c_k \, , \label{BC3bis}
\ee
which will play a key role in the following. Indeed, Eq.~(\ref{BC3bis}) provides boundary conditions for the velocity field at the fluid-solid interface in the presence of diffusiophoresis.  As a consequence of the Onsager-Casimir reciprocal relations~(\ref{OCrr}), the surface current density~(\ref{BC2}) should also involve the diffusiophoretic constants because
\be
L_{k{\rm v}}=- L_{{\rm v}k} = - \frac{1}{k_{\rm B}} \, \lambda \, b_k \, c_k \, ,
\label{L-L}
\ee
expressing microreversibility in the coupling of the slip velocity to the surface current density.

An important issue to note is that the balance equations~(\ref{bal-eq-Gk}) for the excess surface densities are related to the boundary conditions for the concentration fields $c_k^+=c_k$ in solution, which are ruled by the bulk balance equations, as the following reasoning shows. At the fluid-solid interface, we have in Eq.~(\ref{bal-eq-Gk}) that ${\bf j}_k^{+}={\bf j}_k$ is the bulk diffusive current density in the solution and ${\bf j}_k^{-}=0$ in the solid.  If Fick's law holds for species $k$ with bulk diffusion coefficient~$D_k$ in the solution, the current density is given by
\be
{\bf j}_k = c_k{\bf v}- D_k \pmb{\nabla}c_k \, ,
\ee
and we get from Eq.~(\ref{bal-eq-Gk}) that, at the interface,
\be
c_k\, {\bf n}\cdot{\bf v}-D_k \, {\bf n}\cdot \pmb{\nabla}c_k =\sum_{r} \nu_{kr} \, w_{r}^{\rm s}  - \Sigma_k^{\rm s}.
\label{BC1bis}
\ee
Here, $\Sigma_k^{\rm s}$ is the sink into the boundary pool of adsorbate species $k$,
\be
\Sigma_k^{\rm s} \equiv \partial_t\Gamma_k +\pmb{\nabla}_{\bot}\cdot(\Gamma_k{\bf v}^{\rm s}+{\bf j}_k^{\rm s}),
\label{BC1-sink}
\ee
which is expressed in terms of the surface current density
\be
{\bf j}_k^{\rm s}=\frac{\lambda b_k}{k_{\rm B}T}\, c_k\,  {\bf v}_{\rm slip}  -D_{k}^{\rm s}\, \pmb{\nabla}_{\bot}\Gamma_k \, .
\label{BC1-sink-jks}
\ee
Therefore, we have deduced the boundary conditions Eqs.~(\ref{BC3bis}) and~(\ref{BC1bis}) for the velocity and concentrations fields from the phenomenological relations and the balance equations at the interface.

In the following, we shall illustrate the formalism for the simple reaction ${\rm A}\rightleftharpoons{\rm B}$ with stoichiometric coefficients $\nu_{\rm A}=-1$ and $\nu_{\rm B}=+1$.  Its rate is given by
\be
w^{\rm s}= \kappa_+ c_{\rm A}-\kappa_- c_{\rm B} \, ,
\label{ws}
\ee
with rate constants $\kappa_{\pm}$ that are positive on the chemically active surface. Supposing furthermore that the adsorbates can be neglected ($\Gamma_k=0$), the total entropy production rate, including the irreversible processes in the fluid and at the fluid-solid interface, is given by
\bea
&&\frac{1}{k_{\rm B}} \frac{d_{\rm i}S}{dt} = \\
&& \int dV \Big[ \frac{\eta}{2k_{\rm B}T}\left(\pmb{\nabla}{\bf v}+\pmb{\nabla}{\bf v}^{\rm T}\right)^2 + \sum_{k={\rm A},{\rm B}} D_k \frac{(\pmb{\nabla}c_k)^2}{c_k}\Big] +\nonumber\\
&&\int dS \Big[\frac{\lambda}{k_{\rm B}T}\left({\bf v}_{\rm slip}\right)^2 + \left( \kappa_+ \, c_{\rm A} - \kappa_- \, c_{\rm B}  \right)\ln\frac{\kappa_+ \, c_{\rm A}}{\kappa_- \, c_{\rm B}}\Big] \geq 0 \, ,
\nonumber
\eea
where the volume integral extends over the bulk of the fluid and the surface integral over the interface.
The first term is the contribution due to shear viscosity, the second to diffusion, the third to interfacial friction, and the fourth to surface reaction. We notice that the contribution of the diffusiophoretic coupling terms is zero because of cancellation due to the antisymmetry $L_{k{\rm v}}=-L_{{\rm v}k }$.

In the next section, the various coefficients will be determined within the thin-layer approximation.

\section{Thin-layer approximation and boundary conditions}
\label{ThinLayer}

\subsection{Setting up the thin-layer approximation}

\begin{figure}[h]
\centerline{\scalebox{0.35}{\includegraphics{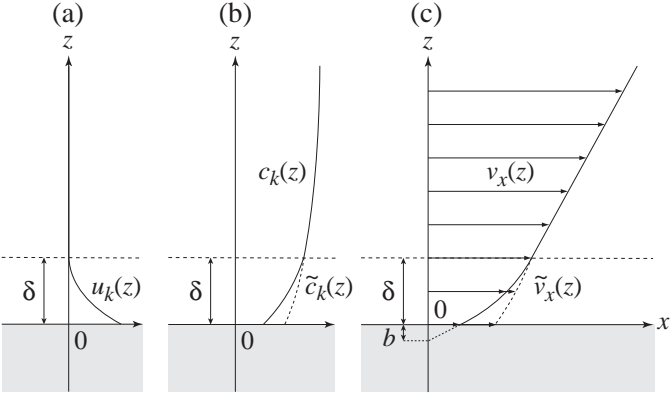}}}
\caption{Schematic representation of the thin-layer approximation: (a) Interaction potential $u_k(z)$ of a solute $k$ with the wall at $z=0$. (b) The actual $c_k(z)$ and effective $\tilde c_k(z)$ concentration profiles.  (c) The actual $v_x(z)$ and effective $\tilde v_x(z)$ velocity profiles.  The width of the boundary layer where the solute interacts with the wall is $\delta$ and $b$ is the slip length.}
\label{fig1}
\end{figure}

We consider a dilute solution of species A and B in a solvent 0 near a solid wall.  The solution is assumed to be incompressible.  As depicted in Fig.~\ref{fig1}, the $z$-axis is perpendicular to the wall, while the $x$-axis is parallel to the wall (as well as the $y$-axis that is not shown for simplicity).  Let $u_k(z)$ denote the interaction potential of solute $k={\rm A}, {\rm B}$ with the wall.  The potentials are taken to vanish beyond their range $z=\delta$, which is assumed to be larger than the molecular size and much smaller than the radius of curvature of the solid interface.

The velocity $\bf v$ and concentration $c_k$  fields obey the following equations:
\bea
&&\rho(\partial_t\,{\bf v} + {\bf v}\cdot\pmb{\nabla}{\bf v})=-\pmb{\nabla}P -\sum_k c_k \pmb{\nabla}u_k + \eta \nabla^2{\bf v} , \qquad \label{NS-eq-actual}\\
&&\pmb{\nabla}\cdot{\bf v} = 0 \, ,\\
&&
\partial_t \,c_k +\pmb{\nabla}\cdot{\bf j}_k = 0 \, , \label{diff-eq-actual}\\
&&{\bf j}_k = c_k{\bf v} -D_k \pmb{\nabla} c_k -\beta D_k c_k\pmb{\nabla} u_k \, ,
\label{jk-actual}
\eea
where $\rho$ is the mass density and $\beta=(k_{\rm B}T)^{-1}$ the inverse temperature. These equations can be rewritten in the following form:
\bea
&&\rho(\partial_t\,{\bf v} + {\bf v}\cdot\pmb{\nabla}{\bf v})=-\pmb{\nabla}\Big(P-k_{\rm B}T\sum_k c_k\Big) \label{NS-eq-actual2}\\
&& \qquad\qquad -k_{\rm B}T \sum_k {\rm e}^{-\beta u_k}\pmb{\nabla}\left({\rm e}^{\beta u_k} c_k\right) + \eta \nabla^2{\bf v}  \, , \nonumber\\
&&\pmb{\nabla}\cdot{\bf v} = 0 \, ,\\
&&
\partial_t \,c_k +\pmb{\nabla}\cdot{\bf j}_k = 0 \, , \label{cons-k-eq-actual2}\\
&&{\bf j}_k = c_k{\bf v} -D_k {\rm e}^{-\beta u_k} \pmb{\nabla}\left( {\rm e}^{\beta u_k}c_k \right)\, . \label{jk-actual2}
\eea

At the wall, the fields satisfy the boundary conditions
\bea
&& v_x(x,0) = b \, \partial_z v_x(x,0) \, , \label{vx-bc}\\
&& v_z(x,0)= 0 \, . \label{vz-bc}
\eea
The slip boundary condition holds for the velocity field $v_x(x,z)$.  Furthermore, the diffusive current density ${\bf j}_k(x,z)$ satisfies the boundary condition
\be
j_{kz}(x,0) = \nu_k \, w(x)
\label{jkz-bc-r1}
\ee
with the surface reaction rate $w(x)$.

The idea behind the thin-layer approximation consists of replacing the actual fields by effective fields $\tilde{\bf v}(x,z)$ and $\tilde c_k(x,z)$ that obey the following equations
\bea
&&\rho(\partial_t\,\tilde{\bf v} + \tilde{\bf v}\cdot\pmb{\nabla}\tilde{\bf v})=-\pmb{\nabla}\tilde P + \eta \nabla^2\tilde{\bf v}\, , \label{NS-eq-eff}\\
&&\pmb{\nabla}\cdot\tilde{\bf v} = 0 \, ,\\
&&
\partial_t \,\tilde c_k +\pmb{\nabla}\cdot\tilde{\bf j}_k = 0 \, , \label{diff-eq-eff}\\
&&\tilde{\bf j}_k = \tilde c_k\tilde{\bf v} -D_k \pmb{\nabla} \tilde c_k \, , \label{tilde-j}
\eea
as if there were no interaction potentials, but with effective boundary conditions that are determined by the analysis developed below.  The effective fields should coincide with the actual fields beyond the boundary layer, i.e., beyond the interaction range of the potentials:
\be
\tilde{\bf v}(x,z) = {\bf v}(x,z) \qquad\mbox{and}\qquad \tilde c_k(x,z)= c_k(x,z)\, ,
\label{matching_eqs}
\ee
if $z\ge \delta$, but they differ within the boundary layer, as schematically depicted in Fig.~\ref{fig1}.

Referring to Eq.~(\ref{decomp-x}), in the limit of an arbitrarily thin boundary layer the actual concentration field $c_k$ governed by Eq.~(\ref{diff-eq-actual}) should correspond to the combination $\tilde c_k\theta^++\Gamma_k\delta^{\rm s}$ in terms of the effective concentration field $\tilde c_k$ and the excess surface density $\Gamma_k$. From Eqs.~(\ref{BC1bis})-(\ref{BC1-sink}) we find that $\Gamma_k$ satisfies
\be
\partial_t\Gamma_k+\partial_x\left(\Gamma_k v_x^{\rm s} +j_{kx}^{\rm s}\right)=\nu_k \, w(x)-\tilde j_{kz}(x,0) \, .
\label{bal_eq_excess}
\ee
Subtracting Eq.~(\ref{diff-eq-eff}) from Eq.~(\ref{diff-eq-actual}), integrating over the thickness $0<z<\delta$ of the boundary layer, and using the boundary condition~(\ref{jkz-bc-r1}), we get
\bea
&&\partial_t\int_0^{\delta}(c_k-\tilde c_k)\, dz \nonumber\\
&&+\, \partial_x\int_0^{\delta} [ c_kv_x-\tilde c_k\tilde v_x-D_k\partial_x(c_k-\tilde c_k)] \, dz \nonumber\\
&&= \nu_k\, w(x)-j_{kz}(x,\delta)+\tilde j_{kz}(x,\delta)-\tilde j_{kz}(x,0) .
\eea
This result may then be compared with Eq.~(\ref{bal_eq_excess}) after using the matching condition $\tilde j_{kz}(x,\delta)=j_{kz}(x,\delta)$, which is a consequence of Eq.~(\ref{matching_eqs}).  We thus obtain the expressions giving the excess surface and current densities:
\bea
&&\Gamma_k(x) = \int_0^{\delta}\left[ c_k(x,z)-\tilde c_k(x,z)\right] dz \, , \label{Gk-thin}\\
&&\Gamma_k v_x^{\rm s} +  j_{kx}^{\rm s} = \int_0^{\delta}\left( c_k v_x -\tilde c_k \tilde v_x \right) dz - D_k \partial_x \Gamma_k\, . \label{jks-thin}\quad
\eea

In order to establish the effective boundary conditions from the actual ones, the fields are assumed to vary much faster in the $z$-direction normal to the surface than in the $x$-direction parallel to the surface.  Accordingly, the partial differential equations can be solved by integrating along the $z$-direction independently of the $x$-direction. Under these conditions the partial derivatives $\partial_x$ of the velocity and concentration fields may be taken to be of order $q_x$, the wavenumber parallel to the surface.  The condition of validity of this approximation is that $ q_x\delta \ll 1$. In addition, $\delta$ is assumed to be much smaller than the characteristic size $H$ of the boundaries: $\delta\ll H$.

At low Reynolds number, Eqs.~(\ref{NS-eq-actual2})-(\ref{jk-actual2}) may be solved for a stationary solution by neglecting the nonlinear term ${\bf v}\cdot\pmb{\nabla}{\bf v}$ in the Navier-Stokes equation.  The components of the current densities~(\ref{jk-actual2}) are given by
\bea
&& j_{kx} = c_kv_x -D_k \partial_x c_k\, , \label{jkx}\\
&& j_{kz} = c_kv_z -D_k {\rm e}^{-\beta u_k} \partial_z\left( {\rm e}^{\beta u_k}c_k \right) , \label{jkz}
\eea
so that Eqs.~(\ref{NS-eq-actual2})-(\ref{cons-k-eq-actual2}) read
\bea
&& \partial_z j_{kz} = -\partial_x(c_kv_x -D_k \partial_x c_k)\,  , \label{eq-jkz} \\
&& \partial_z\left( {\rm e}^{\beta u_k}c_k \right) =- \frac{1}{D_k}\, {\rm e}^{\beta u_k}(j_{kz}-c_k v_z)\,  , \label{eq-ck}\\
&&\partial_z\Big(P-k_{\rm B}T\sum_k c_k\Big)=k_{\rm B}T \sum_k \frac{1}{D_k}\,(j_{kz}-c_k v_z) \nonumber\\
&&\qquad\qquad\qquad\qquad\qquad\qquad + \eta\, (\partial_x^2+\partial_z^2)v_z  \, , \label{eq-P} \\
&& \partial_z^2 v_x = \frac{1}{\eta}\, \partial_x P -\partial_x^2 v_x \, , \label{eq-vx}\\
&&\partial_z v_z = - \partial_x v_x \, . \label{eq-vz}
\eea
The first equation gives the $z$-component of the current density $j_{kz}$, the second the concentration $c_k$, the third the pressure $P$, the fourth the $x$-component of the velocity, and the fifth the $z$-component of the velocity.  Equations can also be deduced for the effective fields ruled by Eqs.~(\ref{NS-eq-eff})-(\ref{tilde-j}), and have the same form as Eqs.~(\ref{eq-jkz})-(\ref{eq-vz}) with the potential set to zero, $u_k=0$.

\subsection{The effective boundary conditions}\label{subsec:effectiveBC}

Equations~(\ref{eq-jkz})-(\ref{eq-vz}) for the actual fields may be successively integrated over $z$ and the resulting integrals can be recursively expanded in powers of $q_x$ and $\delta$ to get expressions in terms of the fields at the boundary $z=0$. Next, a similar integration is performed for the effective fields without specifying their boundary values.  Since the effective fields must coincide with the actual ones if $z\geq \delta$, relationships are obtained between the actual and the effective boundary values.   The details of these calculations are given in Appendix~\ref{AppA}.  As a consequence of  the actual boundary condition~(\ref{vx-bc}), the effective velocity field is found to satisfy
\be
\tilde v_x(x,0) = b \, \partial_z \tilde v_x(x,0) -  \sum_k b_k \, \partial_x \tilde C_k(x) + O(q_x^2\delta), \qquad\label{new-bc}
\ee
expressed in terms of the slip length~(\ref{b}), the diffusiophoretic constants
\be
b_k = \frac{k_{\rm B}T}{\eta} \left( K_k^{(1)} + b\, K_k^{(0)}\right) ,
\label{b_k}
\ee
and the effective concentrations modified by the reaction
\be
\tilde C_k(x) \equiv \tilde c_k(x,0) - \frac{\nu_k}{D_k} \, \varsigma_k \, w(x)\, ,
\label{C-dfn}
\ee
where
\bea
&&\varsigma_k \equiv \frac{1}{K_k^{(1)} + b\, K_k^{(0)}} \biggl[ \Big(K_k^{(2)}-\frac{1}{2}R_k^{(2)}+S_k^{(1)}-K_k^{(1)}R_k^{(0)}\Big)\nonumber\\
&&\qquad\qquad\quad
 +b\left(K_k^{(1)}-R_k^{(1)}+S_k^{(0)}-K_k^{(0)}R_k^{(0)}\right)\biggr]
\label{sigma_k}
\eea
with the quantities
\bea
&& K_k^{(n)} \equiv \int_0^{\delta} dz \, z^n \, \left[ {\rm e}^{-\beta u_k(z)}-1 \right] , \label{K-dfn}\\
&&R_k^{(n)} \equiv \int_0^{\delta} dz \, z^n \, \left[ {\rm e}^{\beta u_k(z)}-1 \right] , \label{R-dfn}\\
&&S_k^{(n)} \equiv \int_0^{\delta} dz \, z^n \, \left[ {\rm e}^{-\beta u_k(z)}-1 \right]  \int_0^{z} dz' \, \left[ {\rm e}^{\beta u_k(z')}-1 \right] , \nonumber\\
&& \label{S-dfn}
\eea
up to corrections $O(q_x^2)$.
The expression~(\ref{b_k}) for the diffusiophoretic constants includes the effect of partial slip and has been obtained in Ref.~\onlinecite{AB06}.  For stick boundary conditions ($b=0$), Eq.~(\ref{b_k}) reduces to the previously known expression.\cite{A89}

We also obtain the boundary condition on the current densities under stationary conditions ($\partial_t\Gamma_k=0$)
\be
\tilde j_{kz}(x,0) = \nu_k \, w(x) - \partial_x(\Gamma_k v_x^{\rm s}+ j_{kx}^{\rm s}) , \label{j-w-djs}
\ee
where the interfacial current densities are calculated using Eq.~(\ref{jks-thin}) to get
\bea
j_{kx}^{\rm s} &=&  \frac{\lambda\, b_k}{k_{\rm B}T} \, \tilde C_k(x) \, \tilde v_x(x,0) \times \left[1+O(\delta/H)\right] \nonumber\\
&& - D_k \, \partial_x\Gamma_k\, +O(q_x\delta^3),
\label{jks}
\eea
in terms of the coefficient of sliding friction $\lambda=\eta/b$, the diffusiophoretic constants~(\ref{b_k}), the effective concentrations~(\ref{C-dfn}), and the diffusion coefficients $D_k$, up to corrections that are negligible under the aforementioned conditions $\delta\ll H$ and $q_x\delta \ll1$.

In summary, substituting the definition~(\ref{C-dfn}) into Eq.~(\ref{new-bc}) and neglecting the corrections,
we obtain the boundary condition for the effective velocity as
\bea
&& \tilde v_x(x,0) = b \, \partial_z \tilde v_x(x,0) \nonumber\\
&&- \sum_k b_k \, \partial_x\Big[\tilde c_k(x,0) - \frac{\nu_k}{D_k} \, \varsigma_k \, w(x)\Big] .
\label{v-bc-fin}
\eea
Next, substituting the expression~(\ref{tilde-j}) for the effective current density into Eq.~(\ref{j-w-djs}), using Eq.~(\ref{jks}) with the definition~(\ref{C-dfn}), and again neglecting the corrections, we get the boundary conditions on the effective concentration fields  as
\be
\tilde c_k(x,0)\tilde v_z(x,0) -D_k \partial_z \tilde c_k(x,0) = \nu_k \, w(x) - \Sigma_k^{\rm s}
\label{c-bc-fin}
\ee
where the sink reads
\bea
&&\Sigma_k^{\rm s} =
\partial_x \biggl\{ \Gamma_k v_x^{\rm s}\nonumber\\
&&+ \frac{\lambda\, b_k}{k_{\rm B}T} \Big[\tilde c_k(x,0) - \frac{\nu_k}{D_k} \, \varsigma_k \, w(x)\Big]\, \tilde v_x(x,0)  - D_k \, \partial_x\Gamma_k\biggr\} . \nonumber\\ &&
\label{S-c-bc-fin}
\eea
Equations~(\ref{c-bc-fin}) and~(\ref{S-c-bc-fin}) are consistent with Eqs.~(\ref{BC1bis}) and~(\ref{BC1-sink}) for stationary conditions.

Comparing Eqs.~(\ref{v-bc-fin}) and (\ref{jks}) with Eqs.~(\ref{BC3bis}) and~(\ref{BC1-sink-jks}), respectively, we see that the role of the concentration fields is played by the effective concentrations~(\ref{C-dfn}) due to the modification by the surface reaction.  In this regard, the effective linear response coefficients are given by
\be
L_{k{\rm v}}=- L_{{\rm v}k} = - \frac{1}{k_{\rm B}} \, \lambda \, b_k \, \tilde C_k(x) \, .
\label{L-L-eff}
\ee

Furthermore, we notice that, if the reaction rate is a linear combination of the concentrations, the first boundary condition~(\ref{v-bc-fin}) can be written as
\be
\tilde v_x(x,0) = b \, \partial_z \tilde v_x(x,0) - \sum_k \tilde b_k \, \partial_x \tilde c_k(x,0)
\label{v-bc-eff-bk}
\ee
in terms of renormalized diffusiophoretic constants $\tilde b_k$. The results for stick a boundary
condition ($b=0$) are summarized in Appendix~\ref{AppB}.

These results show that the diffusiophoretic effect is described as a coupling between two interfacial irreversible processes, namely, the sliding friction and transport by concentration gradients.

\subsection{Specialization to the ${\rm A}\rightleftharpoons{\rm B}$ reaction}

For the ${\rm A}\rightleftharpoons{\rm B}$ reaction, the rate is given by
\be
w(x) = \kappa_+^{\rm s}\, c_{\rm A}(x,0) - \kappa_-^{\rm s}\, c_{\rm B}(x,0) \, ,
\label{wAB}
\ee
where the rate constants $\kappa_{\pm}^{\rm s}$ are defined at the catalytic surface.

Referring to Appendix~\ref{AppA}, the reaction rate~(\ref{wAB}) may also be expressed in terms of the effective concentrations by inverting Eqs.~(\ref{tilde_c-c-w}), giving
\bea
&&w(x) = \frac{\kappa_+^{\rm s}}{\Delta}\, {\rm e}^{-\beta u_{\rm A}(0)}\, \tilde c_{\rm A}(x,0) - \frac{\kappa_-^{\rm s}}{\Delta}\, {\rm e}^{-\beta u_{\rm B}(0)} \,\tilde c_{\rm B}(x,0)\nonumber\\
&& \qquad\qquad +O(q_x) \, ,
\label{wAB-correct}
\eea
where
\begin{equation}
\Delta=1+ \frac{\kappa_+^{\rm s}}{D_{\rm A}} \, {\rm e}^{-\beta u_{\rm A}(0)} R_{\rm A}^{(0)}
+ \frac{\kappa_-^{\rm s}}{D_{\rm B}}\, {\rm e}^{-\beta u_{\rm B}(0)} R_{\rm B}^{(0)},
\end{equation}
so that the reaction rate can be rewritten as
\be
w(x) = \kappa_+\,  \tilde c_{\rm A}(x,0) - \kappa_-\,\tilde c_{\rm B}(x,0)
\label{wAB-redfn}
\ee
with renormalized rate constants $\kappa_{\pm}$. As expected, the rate constants are modified by Boltzmann factors corresponding to the interaction energies at the surface.  Moreover, the denominator $\Delta$ includes corrections depending on the quantities~(\ref{R-dfn}).

For the ${\rm A}\rightleftharpoons{\rm B}$ reaction~(\ref{wAB}), the slip velocity~(\ref{v-bc-eff-bk}) becomes
\be
\tilde v_x(x,0) = b \, \partial_z \tilde v_x(x,0) -  \tilde b_{\rm A} \, \partial_x \tilde c_{\rm A}(x,0) -  \tilde b_{\rm B} \, \partial_x \tilde c_{\rm B}(x,0)
\label{v-bc-eff-bk-2}
\ee
with the renormalized diffusiophoretic coefficients:
\bea
&& \tilde b_{\rm A} = b_{\rm A}\left( 1 +  \frac{\varsigma_{\rm A}}{D_{\rm A}}\, \kappa_+\right) - b_{\rm B} \, \frac{\varsigma_{\rm B}}{D_{\rm B}}\, \kappa_+ \, , \label{renorm-bA}\\
&& \tilde b_{\rm B} = b_{\rm B} \left( 1 + \frac{\varsigma_{\rm B}}{D_{\rm B}}\, \kappa_-\right)- b_{\rm A} \, \frac{\varsigma_{\rm A}}{D_{\rm A}}\, \kappa_- \, , \label{renorm-bB}
\eea
$\varsigma_{\rm A}$ and $\varsigma_{\rm B}$ being given by Eq.~(\ref{sigma_k}).

\subsection{Estimates of the corrections}

In order to estimate the relative importance of these corrections, we may consider the simple square-well potentials
\be
u_k(z) = \left\{
\begin{array}{ll}
u_k(0) \, , & \quad \mbox{if} \quad 0<z<\delta \, , \\
0 \, , & \quad \mbox{if} \quad \delta \le z \, .
\end{array}
\right.
\ee
In this case, the constants take the following values:
\bea
&&K_k^{(n)} =\phi_k^{-} \frac{\delta^{n+1}}{n+1} , \quad
R_k^{(n)} = \phi_k^{+}\frac{\delta^{n+1}}{n+1} , \quad\mbox{and} \nonumber\\
&&S_k^{(n)} =\phi_k^{+}\phi_k^{-}\frac{\delta^{n+2}}{n+2}  , \ \  \mbox{with}\ \ \phi_k^{\pm}\equiv {\rm e}^{\pm\beta u_k(0)}-1 .
 \label{KRSn-estim}
\eea

Taking the no-slip boundary condition $b=0$, we have that the bare diffusiophoretic constant~(\ref{b_k}) is equal to
\be
b_k = \frac{k_{\rm B}T}{\eta} \, K_k^{(1)} = \frac{k_{\rm B}T}{\eta} \, \phi_k^{-} \frac{\delta^{2}}{2} \, ,
\ee
the constant~(\ref{sigma_k}) to $\varsigma_k = \delta$,  and $\Delta$ that enters the rate~(\ref{wAB-correct}) becomes

\be
\Delta = 1- \frac{\kappa_+^{\rm s}}{D_{\rm A}} \, \phi_{\rm A}^{-} \, \delta  -  \frac{\kappa_-^{\rm s}}{D_{\rm B}} \, \phi_{\rm B}^{-} \, \delta \, .
\label{wAB-correct-2}
\ee
We notice that $\varsigma_k = \delta$ even if $b\neq 0$ in this model.

Denoting $H$ the characteristic size of the boundaries, we introduce the Damk\"ohler number
\be
{\rm Da} = H \, \frac{\kappa_{\pm}^{\rm s}}{D_k} \, ,
\ee
characterizing the diffusive control of the reaction.
The reaction proceeds in the reaction-limited regime if ${\rm Da}\ll 1$, or in the diffusion-limited one if ${\rm Da}\gg 1$.  Accordingly, the relative modifications of the renormalized diffusiophoretic constants~(\ref{renorm-bA}) and~(\ref{renorm-bB}) behave as
\be
\Big\vert \frac{\tilde b_k-b_k}{b_k}\Big\vert \sim {\rm Da} \, \frac{\delta}{H} ,
\ee
because $\varsigma_k = \delta$.  Since the range $\delta$ of the interaction potentials $u_k(z)$ is typically of the order of nanometers in solutions of neutral species, we have that $\delta/H\sim 10^{-3}$ for a micrometric size $H$.  Therefore, the modifications of the diffusiophoretic constants remain negligible in the reaction-limited regime where
${\rm Da}\ll 1$, but can lead to significant effects in the diffusion-limited regime for ${\rm Da}\sim 10^3$ or larger.

Concerning the rate constants, an additional effect comes from the Boltzmann factors ${\rm e}^{-\beta u_k(0)}$
in $\Delta$ in Eq.~(\ref{wAB-correct-2}).  If $u_k(0)\sim k_{\rm B}T$, the Boltzmann factors are of order unity.  In this case, the corrections in the denominator are also of order ${\rm Da}\,\delta/H$, which can become significant if ${\rm Da}\gtrsim 10^3$.

\section{Numerical examples}
\label{Ex}

In order to test the predictions of the thin-layer approximation, we consider the Poiseuille flow of a solution between two parallel planes with identical properties and separated by a distance $H$.  The Navier-Stokes and advection-diffusion equations are numerically integrated by the standard staggered-grid method\cite{GDN98} in a two-dimensional domain with $0<x<L$ and $0<z<H/2$ using the mirror symmetry with respect to the plane $z=H/2$. In the two examples below, the actual fields are compared to the effective fields.

On one hand, the numerical integration of the Navier-Stokes and advection-diffusion equations~(\ref{NS-eq-actual})-(\ref{jk-actual}) for the actual fields is performed with the inflow-outflow boundary conditions
\bea
&& v_x(0,z) = \frac{g}{2\eta} \, (z^2-Hz-Hb) , \qquad \partial_x v_x(L,z)=0 , \label{num-bc-vx}\\
&& v_z(0,z) = 0 , \qquad\qquad\qquad\qquad\qquad \partial_x v_z(L,z)=0 , \label{num-bc-vz}\\
&& c_k(0,z)=c_{k,{\rm in}}\, {\rm e}^{-\beta u_k(z)} , \quad c_k(L,z)=c_{k,{\rm out}}\, {\rm e}^{-\beta u_k(z)} , \qquad
\eea
where $g=\partial_x P$, together with the boundary conditions~(\ref{vx-bc})-(\ref{jkz-bc-r1}) on the planes.  The mass density takes the value $\rho=1$, the viscosity $\eta=0.1$, the slip length $b=0.1$, the pressure gradient $g=-0.1$, $H=2$, and $L=1$.  Moreover, the diffusion coefficients are $D_{\rm A}=D_{\rm B}=1$, the inverse temperature $\beta=2$, and the interaction potentials of the solutes with the wall are given by
\be
u_{\rm A}(z) =
\left\{
\begin{array}{ll}
10\times (\delta-z)^2 & \mbox{if} \quad 0<z<\delta , \\
0  &\mbox{if} \quad \delta<z ,
\end{array}
\right.
\label{uA}
\ee
with $\delta=0.25$, and $u_{\rm B}=0$, so that the diffusiophoretic effect is only due to the species A.  The unit of length is taken as $H/2$.  The chosen parameter values correspond to a flow of Reynolds number ${\rm Re}=\rho H v_x(0,H/2)/\eta=12$. The ratio $\delta/H=0.125$ is taken large enough in order to visualize the boundary layer in the numerical integration.

On the other hand, the integration of Eqs.~(\ref{NS-eq-eff})-(\ref{tilde-j}) for the effective fields is carried out with the same inflow-outflow boundary conditions as in Eqs.~(\ref{num-bc-vx}) and~(\ref{num-bc-vz}), but $\tilde c_k(0,z)=c_{k,{\rm in}}$ and $\tilde c_k(L,z)=c_{k,{\rm out}}$, together with the boundary conditions~(\ref{v-bc-fin}) and~(\ref{c-bc-fin}) in the approximation $\Sigma_k^{\rm s}=0$ on the planes.

\subsection{Example without reaction}

\begin{figure}[h]
\centerline{\scalebox{0.45}{\includegraphics{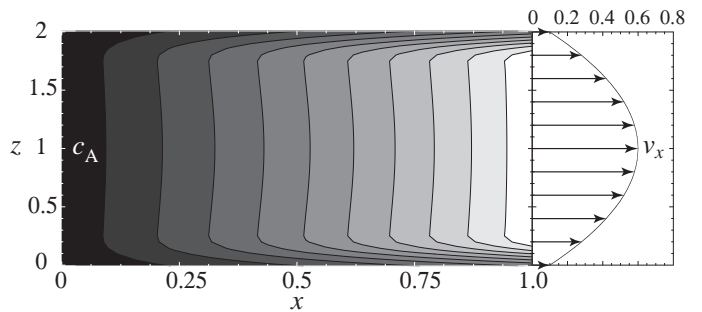}}}
\caption{Poiseuille flow of a solution with solute A between two planes without reaction for the boundary conditions $c_{\rm A,in}=0$, $c_{\rm A,out}=5$, and other parameter values specified in the text: The stationary concentration field $c_{\rm A}(x,z)$ in the two-dimensional domain $(x,z)$ and the velocity profile $v_x(L,z)$ versus $z$ are shown.  The grid size is $\Delta x=\Delta z=0.05$.}
\label{fig2}
\end{figure}

In this first example, the planes are chemically inactive and the solution contains only the solute species A, in order to test the diffusiophoretic boundary conditions.  At the boundaries $x=0$ and $x=L$, the concentrations are taken as $c_{\rm A,in}=0$, $c_{\rm A,out}=5$, and $c_{\rm B,in}=c_{\rm B,out}=0$, so that $c_{\rm B}(x,z)=0$. The stationary concentration field $c_{\rm A}(x,z)$ is shown in Fig.~\ref{fig2}, together with the stationary profile $v_x(L,z)$ of the outflow.  Since the  interaction potential~(\ref{uA}) is repulsive, there is a deficit of species A in the boundary layer of width $\delta=0.25$ near the walls.

\begin{figure}[h]
\centerline{\scalebox{0.5}{\includegraphics{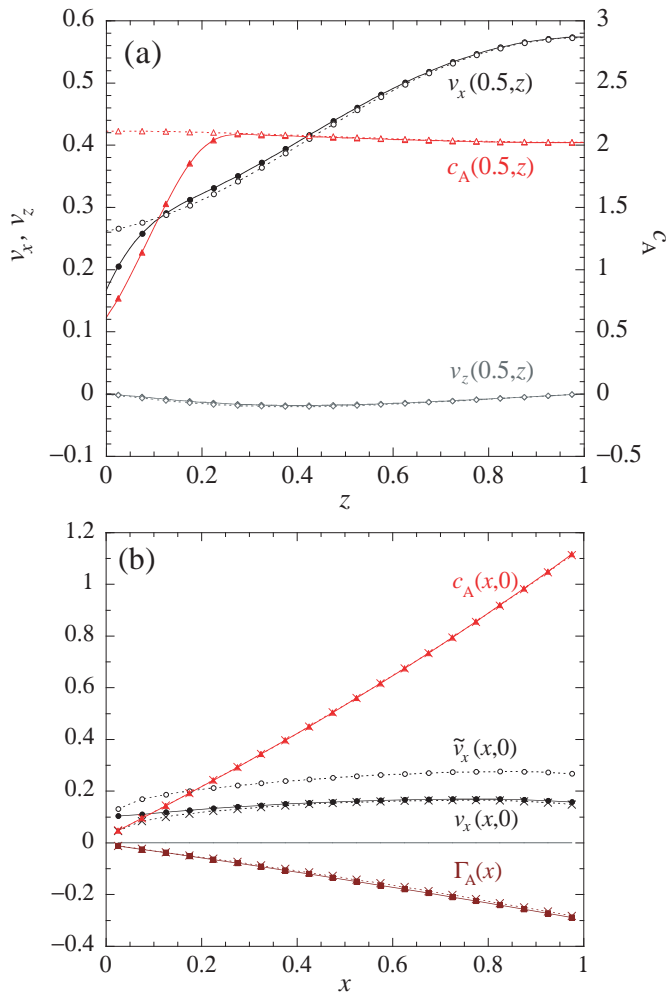}}}
\caption{Poiseuille flow of a solution with solute A between two planes without reaction for the boundary conditions $c_{\rm A,in}=0$, $c_{\rm A,out}=5$, and other parameter values specified in the text:  Panel~(a) shows the two components of the velocity field as well as the concentration field $c_{\rm A}$ along the $z$-direction at $x=0.5$.  The actual fields are depicted by solid lines and filled symbols, and the effective fields by dashed lines and open symbols.  Panel~(b) shows the velocity field $v_x$ as well as the concentration field $c_{\rm A}$ along the $x$-direction at the wall $z=0$.  The crosses depict the predictions of the thin-layer approximation. The grid size is $\Delta x=\Delta z=0.05$.}
\label{fig3}
\end{figure}

Figure~\ref{fig3} compares the actual and effective fields along the $z$-direction in the middle of the domain at $x=0.5$ (Fig.~\ref{fig3}a), and along the $x$-direction at the wall $z=0$ (Fig.~\ref{fig3}b).

On the walls, the boundary conditions of the effective fields are given by Eqs.~(\ref{v-bc-fin}) and~(\ref{c-bc-fin}) with $w(x)=0$ since there is no reaction in this example.  In Eq.~(\ref{v-bc-fin}), the diffusiophoretic constant for species A is calculated with Eqs.~(\ref{b_k}) and~(\ref{K-dfn}), giving the value $b_{\rm A}=-0.06313$.  In Eq.~(\ref{c-bc-fin}), the approximation $\Sigma_{\rm A}^{\rm s}=0$ is taken, supposing that there is no sink into the boundary layer.

In Fig.~\ref{fig3}a, we see the profiles at $x=0.5$ in the $z$-direction for the two components of the velocity field and the concentration field of species A.  The agreement between the actual and effective fields away from the wall shows that the thin-layer approximation provides suitable boundary conditions on the effective fields.  Near the wall, we observe differences between the actual and effective fields, as schematically represented in Fig.~\ref{fig1}.  The actual concentration is essentially related to the effective one by $c_{\rm A}(x,z)={\rm e}^{-\beta u_{\rm A}(z)}\tilde c_{\rm A}(x,z)$, as expected by Eqs.~(\ref{ck-actual}), (\ref{c-tilde}), and~(\ref{ck0-actual-eff}).  The actual velocity field $v_x(x,z)$ is lower than the effective one $\tilde v_x(x,z)$ near the wall, which is the effect of diffusiophoresis.  Since the slip length is positive ($b=0.1$), the actual velocity field is not vanishing at the wall where $v_x(x,0)=b\partial_z v_x(x,0)$.  However, the effective velocity field at the wall is larger because of the diffusiophoretic contribution:  $\tilde v_x(x,0)=b\partial_z \tilde v_x(x,0)-b_{\rm A}\partial_x \tilde c_{\rm A}(x,0)$.  In addition, the $z$-component of the velocity field is very small.

In Fig.~\ref{fig3}b, the boundary values of the concentration field $c_{\rm A}$ and velocity fields $v_x$ and $\tilde v_x$ at the wall $z=0$ are shown along the $x$-direction, together with the excess surface density $\Gamma_{\rm A}(x)$.  The crosses depict the predictions of the thin-layer approximation for the different quantities.  For the concentration $c_{\rm A}(x,0)$, the agreement between the filled triangles and the crosses is the verification of the relation $c_{\rm A}(x,0)={\rm e}^{-\beta u_{\rm A}(0)}\tilde c_{\rm A}(x,0)$ predicted by Eq.~(\ref{ck0-actual-eff}) with $w=0$.  For the velocity field $v_x(x,0)$, the agreement between the filled circles and the crosses is the verification of the relation $v_x(x,0)=\tilde v_x(x,0)+(k_{\rm B}T/\eta) K_{\rm A}^{(1)}\partial_x \tilde c_{\rm A}(x,0)$ predicted by Eq.~(\ref{vx-vx-tilde}) with the value $K_{\rm A}^{(1)}=-0.00519$ given by Eq.~(\ref{K-dfn}).  The deviations near $x=0$ and $x=1$ are the perturbations of inflow and outflow.  The excess surface density $\Gamma_{\rm A}(x)$ calculated by Eq.~(\ref{Gk-thin}) is also in good agreement with the prediction given by Eq.~(\ref{Gamma_k}) with $K_{\rm A}^{(0)}=-0.07439$ and $w=0$.  The excess surface density $\Gamma_{\rm A}(x)$ is negative because of the repulsive interaction of species~A with the wall.

The agreement with the approximation that neglects the sink, $\Sigma_{\rm A}^{\rm s}=0$, means that its contribution is much smaller than the other effects.

\subsection{Example with reaction}

In this second example, the system is the same as above except that the reaction ${\rm A}\rightleftharpoons{\rm B}$ is catalyzed on the parallel planes.  Therefore, the solution contains the solute species A and B. Our aim is to test the effects of the surface reaction~(\ref{wAB}) with $\kappa_+^{\rm s}=\kappa_-^{\rm s}=5$.  The boundary conditions on the actual velocity field are the same as before, but we now have that $c_{\rm A,in}=0$, $c_{\rm A,out}=5$, $c_{\rm B,in}=2$, and $c_{\rm B,out}=0$, corresponding to a Damk\"ohler number equal to ${\rm Da}=H\kappa_{\pm}^{\rm s}/D_k=10$.

\begin{figure}[h]
\centerline{\scalebox{0.5}{\includegraphics{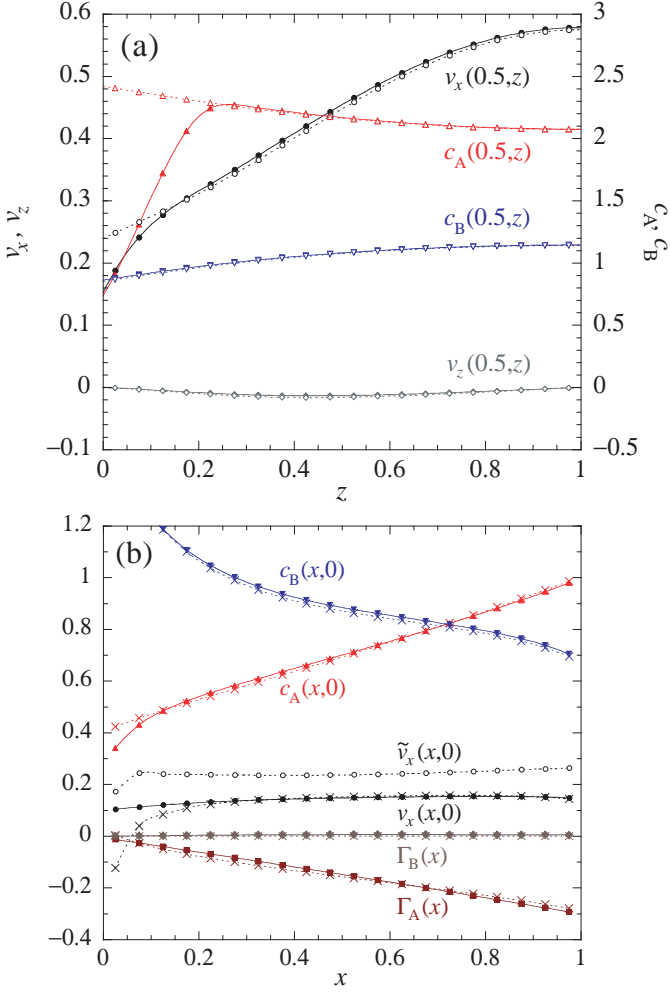}}}
\caption{Poiseuille flow of a solution with solute A between two catalytic planes for the boundary conditions $c_{\rm A,in}=0$, $c_{\rm A,out}=5$, $c_{\rm B,in}=2$, $c_{\rm B,out}=0$, and the other ones specified in the text:  Panel~(a) shows the two components of the velocity field as well as the concentration field $c_{\rm A}$ along the $z$-direction at $x=0.5$.  The actual fields are depicted by solid lines and filled symbols, and the effective fields by dashed lines and open symbols.  Panel~(b) shows the velocity field $v_x$ as well as the concentration field $c_{\rm A}$ along the $x$-direction at the wall $z=0$.  The crosses depict the predictions of the thin-layer approximation. }
\label{fig4}
\end{figure}

The different actual and effective fields are compared in Fig.~\ref{fig4} to test the validity of the thin-layer approximation.  According to Eq.~(\ref{wAB-correct}), the renormalized rate constants appearing in Eq.~(\ref{wAB-redfn}) take the following values:
\be
\kappa_+ = 1.16761 \, , \qquad \kappa_-= 4.07537 \, ,
\label{renorm-rate-csts}
\ee
for the present example where ${\rm e}^{-\beta u_{\rm A}(0)}=0.28651$ and $R_{\rm A}^{(0)}=0.15838$.  The diffusiophoretic constant of species~A is the same as before, $b_{\rm A}=-0.06313$, but the reaction modifies the effective concentration according to Eq.~(\ref{C-dfn}) with the coefficient $\varsigma_{\rm A}=0.13231$ given by Eq.~(\ref{sigma_k}).  Consequently, the renormalized diffusiophoretic constants~(\ref{renorm-bA}) and~(\ref{renorm-bB}) in Eq.~(\ref{v-bc-eff-bk-2}) are given by
\be
\tilde b_{\rm A}= -0.07288 \, , \qquad \tilde b_{\rm B} =  0.03404\, .
\ee

Figure~\ref{fig4}a compares the profiles of the actual and effective fields $v_x$, $v_z$, $c_{\rm A}$, and $c_{\rm B}$, along the $z$-direction at $x=0.5$.  As before, there is good agreement between the actual and effective fields away from the wall at $z=0$.  Since there is a repulsive interaction of species A with the wall, the actual concentration $c_{\rm A}$ is smaller near the wall than the effective concentration $\tilde c_{\rm A}$, essentially by the Boltzmann factor ${\rm e}^{-\beta u_{\rm A}(z)}$.  We also see the diffusiophoretic effect on the velocity field near the wall at $z=0$, where the actual field $v_x(0.5,z)$ is smaller than the effective field $\tilde v_x(0.5,z)$ because of the diffusiophoretic slip given by Eq.~(\ref{v-bc-eff-bk}).  One of the main points is that the agreement between the actual and effective concentration fields is a direct consequence of the renormalized rate constants~(\ref{renorm-rate-csts}).  The reduction of the forward rate constant $\kappa_+$ with respect to the unrenormalized value $\kappa_+^{\rm s}=5$ implies, on the one hand, that the concentration of species A is
larger than in the absence of the reaction, as we see comparing Figs.~\ref{fig3} and~\ref{fig4}, and on the other hand, that the concentration of species B is lower than it would be if the rate constants had their unrenormalized values.  The agreement between the actual and effective concentration fields $c_{\rm A}$ and $c_{\rm B}$ is thus directly determined by the renormalization~(\ref{wAB-correct}) of the rate constants.  As before, the component $v_z$ of the velocity field is very small.

Figure~\ref{fig4}b shows the boundary values of the concentration and velocity fields at the wall $z=0$ along the $x$-direction, together with the excess surface densities $\Gamma_{\rm A}$ and $\Gamma_{\rm B}$.  The actual quantities are compared with the predictions of the thin-layer approximation (crosses).  There is good agreement inside the domain with deviations near the entrance and exit because of perturbations due to the inflow and outflow, as before.  For the concentration fields, the predictions of Eq.~(\ref{ck0-actual-eff}) are that
\be
c_{\rm A}(x,0)= {\rm e}^{-\beta u_{\rm A}(0)} \biggl[ \tilde c_{\rm A}(x,0)-\frac{R_{\rm A}^{(0)}}{D_{\rm A}}\, w(x)\biggr]
\label{cA-predict}
\ee
and $c_{\rm B}(x,0)=\tilde c_{\rm B}(x,0)$, which are in good agreement with the values of the actual fields as seen in Fig.~\ref{fig4}b.  For the velocity field $v_x(x,0)$, the prediction of Eq.~(\ref{vx-vx-tilde}) is that
\be
v_x(x,0)=\tilde v_x(x,0)+\frac{k_{\rm B}T}{\eta}\partial_x \biggl[ K_{\rm A}^{(1)}\tilde c_{\rm A}(x,0)+ \frac{\xi_{\rm A}}{D_{\rm A}} \, w(x) \biggr]
\label{vx0-Ex2}
\ee
with the constants $K_{\rm A}^{(1)}=-0.00519$ and
\be
\xi_{\rm A}=K_{\rm A}^{(2)}-\frac{1}{2}\, R_{\rm A}^{(2)}+S_{\rm A}^{(1)}-K_{\rm A}^{(1)}R_{\rm A}^{(0)}= -7.7320\times 10^{-4} \, .
\ee
As seen in Fig.~\ref{fig4}b, the actual field $v_x(x,0)$ (filled circles) is in agreement with this prediction (crosses).  In Eq.~(\ref{vx0-Ex2}), the term due to the reaction $w(x)$ contributes about 25\% of the diffusiophoretic slip.  Also, the prediction of Eq.~(\ref{Gamma_k}) for the excess surface densities is given by
\be
\Gamma_{\rm A}(x) = K_{\rm A}^{(0)} \tilde c_{\rm A}(x,0)+ \frac{\varepsilon_{\rm A}}{D_{\rm A}} \, w(x)
\ee
with the constants $K_{\rm A}^{(0)}=-0.07439$ and
\be
\varepsilon_{\rm A}=K_{\rm A}^{(1)}-R_{\rm A}^{(1)}+S_{\rm A}^{(0)}-K_{\rm A}^{(0)}R_{\rm A}^{(0)}= -8.9718\times 10^{-3} \, ,
\ee
while $\Gamma_{\rm B}(x) =0$.  We observe in Fig.~\ref{fig4}b the agreement between these predictions (crosses) and the excess surface densities $\Gamma_{\rm A}(x)$ and $\Gamma_{\rm B}(x)$ calculated by their definition~(\ref{Gk-thin}) (filled symbols).

Given that there is no fitting parameter in the comparison, the observed agreement brings numerical support to the validity of the boundary conditions derived for the effective fields within the thin-layer approximation and their modifications by the surface reaction.

\section{Conclusion and perspectives}
\label{Conclusion}

Starting from nonequilibrium interfacial thermodynamics, we derived the boundary conditions at a fluid-solid interface for the fluid velocity and concentration fields that are relevant for diffusiophoresis in the presence of surface reactions.  These conditions play a key role in the self-propulsion of active nano- or micrometric particles made of catalytic solid material, as well as the operation of chemically self-powered micropumps.\cite{SSK15}  Diffusiophoresis is the coupling of solute concentration gradients to the tangential components of the pressure tensor at the interface.  In virtue of microreversibility, nonequilibrium interfacial thermodynamics shows that diffusiophoresis has a reciprocal effect, which couples the interfacial solute current density to the slip velocity (or its gradient in the case of stick boundary conditions).  Since diffusiophoresis and its reciprocal effect couple quantities with opposite parity under time reversal, the corresponding linear response coefficients obey antisymmetric Onsager-Casimir reciprocal relations.  The situation is the same as that for the coefficients characterizing thermal slip in thermophoresis.\cite{W67,BAM76}

Solute concentrations are modified near the interface in the presence of surface reactions.  In order to investigate the consequences of these modifications, the interaction of the solute species with a solid surface can be described by interaction potentials of finite range, defining the thickness of the boundary layer between the fluid and the surface.  The sliding friction between the fluid and the solid surface can be taken into account by partial slip boundary conditions and the associated slip length.  Integrating the Navier-Stokes and advection-diffusion equations in the direction normal to the surface, the diffusiophoretic coupling constants can be obtained within the framework of the thin-layer approximation.\cite{A89,AP91}  This method establishes a matching between actual and effective fields in the limit where the boundary layer is arbitrarily thin with respect to the interfacial radius of curvature.  In this framework, we have shown that surface reactions may change the diffusiophoretic coupling, as well as its reciprocal effect.  Accordingly, the boundary conditions on the velocity and concentration fields can be modified at the reactive interface.  The modifications may have a significant impact if the reaction rate constants and the interaction potential of reactive species with the surface are large.  Analytical formulas describing these effects have been derived in the thin-layer approximation, which confirms the antisymmetric Onsager-Casimir reciprocal relations.  These results are tested numerically for the Poiseuille flow of a solution between to parallel planes that are either chemically inactive or catalytic, showing agreement with the predictions of the thin-layer approximation.

Several issues have been left open in this study.  We considered a fluid-solid interface and supposed that the boundary layer was in quasiequilibrium with the bulk solution.  Moreover, the partial slip and the reactive boundary conditions were taken at the solid surface.  Beyond such situations, there are fluid-fluid interfaces with interfacial viscosities and the transport processes to and from the interface may have finite characteristic time scales so that the boundary layer could be out of equilibrium with respect to the bulk solution.  Furthermore, the interfacial sliding friction as well as the surface reaction also involve the interaction of the solute and solvent species with the interface.  It should be pointed out that the interaction potentials describing each one of these effects have their own characteristic length scales, which may have different relative magnitudes depending on the system of interest.  In this respect, we can envisage systems where sliding friction, diffusiophoresis, and surface reaction could be coupled in a way that uses assumptions that differ from those in this paper.  We notice that molecular dynamics simulations should be used if several of the characteristic length scales are of molecular size.  Moreover, nonlinear surface reactions with adsorbate species diffusing along the surface may also be studied, instead of the simple reaction ${\rm A}\rightleftharpoons{\rm B}$. These considerations can be extended to treat the electrophoretic and thermophoretic mechanisms.

\section*{Acknowledgments}

The Authors thank Patrick Grosfils and Mu-Jie Huang for fruitful discussions.
Financial support from the International Solvay Institutes for Physics and Chemistry, the Universit\'e libre de Bruxelles (ULB), the Fonds de la Recherche Scientifique~-~FNRS under the Grant PDR~T.0094.16 for the project ``SYMSTATPHYS", and the Natural Sciences and Engineering Research Council of Canada is acknowledged.


\appendix
\section{Calculations of the thin-layer approximation}
\label{AppA}

In this Appendix we provide additional details concerning the calculations leading to the effective boundary condition in Sec.~\ref{subsec:effectiveBC}. The computation starts by successively integrating Eqs.~(\ref{eq-jkz})-(\ref{eq-vz}) for the actual fields over $z$.  After recursive substitutions, the fields are given as expansions in powers of $q_x$ involving the fields at the boundary $z=0$ (where $q_x$ denotes the gradients in the velocity and concentration fields).

First, we note that integrating Eq.~(\ref{eq-vz}) with the boundary condition~(\ref{vz-bc}) shows that $v_z=O(q_xz)$, whereupon $\partial_x^2 v_z=O(q_x^3z)$ in Eq.~(\ref{eq-P}) and $\partial_x^2 v_x=O(q_x^2)$ in Eq.~(\ref{eq-vx}).

With the boundary condition~(\ref{jkz-bc-r1}), the integration of Eq.~(\ref{eq-jkz}) over $z$ leads to
\begin{eqnarray*}
&& j_{kz}(x,z) =\nu_k \, w(x)\nonumber\\
&& -\partial_x\left[c_k(x,0)\, v_x(x,0)\right] {\rm e}^{\beta u_k(0)} \int_0^z dz' {\rm e}^{-\beta u_k(z')} \nonumber\\
&&-\partial_x\left[c_k(x,0)\, \partial_z v_x(x,0)\right] {\rm e}^{\beta u_k(0)} \int_0^z dz' \, z' \, {\rm e}^{-\beta u_k(z')} \nonumber\\
&&-\frac{1}{2\eta}\, \partial_x\left[c_k(x,0)\, \partial_xP(x,0)\right] {\rm e}^{\beta u_k(0)} \int_0^z dz' \, z'^2 \, {\rm e}^{-\beta u_k(z')} \nonumber\\
&&+ \frac{\nu_k}{D_k} \, \partial_x\left[w(x)\, v_x(x,0)\right] \nonumber\\
&&\qquad\times \int_0^z dz' {\rm e}^{-\beta u_k(z')} \int_0^{z'} dz'' \, {\rm e}^{\beta u_k(z'')} \nonumber\\
\end{eqnarray*}
\bea
&&+ \frac{\nu_k}{D_k} \, \partial_x\left[w(x)\, \partial_zv_x(x,0)\right]\nonumber\\
&&\qquad\times \int_0^z dz' \, z' \, {\rm e}^{-\beta u_k(z')} \int_0^{z'} dz'' \, {\rm e}^{\beta u_k(z'')} \nonumber\\
&&+ \frac{\nu_k}{2\eta D_k} \, \partial_x\left[w(x)\, \partial_xP(x,0)\right]\nonumber\\
&&\qquad\times \int_0^z dz' \, z'^2 \, {\rm e}^{-\beta u_k(z')} \int_0^{z'} dz'' \, {\rm e}^{\beta u_k(z'')} \nonumber\\
&&+ O(q_x^2z)  .
\label{A1}
\eea

Integrating Eq.~(\ref{eq-ck}) over $z$, the concentrations are given by
\bea
\label{ck-actual}
&&c_k(x,z) = {\rm e}^{-\beta u_k(z)}  {\rm e}^{\beta u_k(0)} c_k(x,0)
\nonumber\\
&&- \frac{\nu_k}{D_k} \, w(x) \, {\rm e}^{-\beta u_k(z)} \int_0^z dz'  \, {\rm e}^{\beta u_k(z')} + O(q_xz^2) .
\label{A2}
\eea

The integration of Eq.~(\ref{eq-P}) gives the pressure:
\bea
&& P(x,z) = P(x,0) + k_{\rm B}T \sum_k c_k(x,0)\Big[{\rm e}^{-\beta u_k(z)}{\rm e}^{\beta u_k(0)}-1 \Big]
\nonumber\\
&& - k_{\rm B}T\, w(x) \sum_k\frac{\nu_k}{D_k}\Big[ {\rm e}^{-\beta u_k(z)}\int_0^z dz'{\rm e}^{\beta u_k(z')}-z \Big] + O(q_xz) .
\nonumber\\
&&
\label{A3}
\eea

Next, the actual velocity field is given by integrating Eqs.~(\ref{eq-vx}) and~(\ref{eq-vz}):
\bea
&& v_x(x,z) = v_x(x,0) + z \, \partial_z v_x(x,0) + \frac{z^2}{2\eta}\, \partial_x P(x,0) \nonumber\\
&&+ \frac{k_{\rm B}T}{\eta} \sum_k \partial_x c_k(x,0) \nonumber\\
&&\times\int_0^z dz' \int_0^{z'} dz'' \left[ {\rm e}^{-\beta u_k(z'')} {\rm e}^{\beta u_k(0)} - 1\right] \nonumber\\
&&- \frac{k_{\rm B}T}{\eta} \, \partial_x w(x) \sum_k \frac{\nu_k}{D_k} \int_0^z dz' \int_0^{z'} dz''
\nonumber\\
&&\times \biggl[ {\rm e}^{-\beta u_k(z'')}  \int_0^{z''} dz''' \, {\rm e}^{\beta u_k(z''')} -  z'' \biggr]
\nonumber\\
&&+ O(q_x^2z^2) ,
\label{A4}
\eea
\bea
&& v_z(x,z) =  -z \, \partial_x v_x(x,0) - \frac{z^2}{2}\, \partial_x\partial_z v_x(x,0)\nonumber\\
&&\qquad\qquad\qquad - \frac{z^3}{6\eta} \,
\partial_x^2 P(x,0) + O(q_x^2z^2) .
\label{A5}
\eea

A similar integration is performed for the effective fields without specifying their boundary values to get the effective current density of species $k$
\begin{eqnarray*}
&&\tilde j_{kz}(x,z) = \tilde j_{kz}(x,0) - z\, \partial_x\left[ \tilde c_k(x,0)\,  \tilde v_x(x,0)\right] \nonumber\\
&&- \frac{z^2}{2}\, \partial_x\left[ \tilde c_k(x,0)\, \partial_z  \tilde v_x(x,0)\right] \nonumber\\
&&- \frac{z^3}{6\eta}\, \partial_x\left[ \tilde c_k(x,0)\, \partial_x  \tilde P(x,0)\right] \nonumber\\
&&+ \frac{z^2}{2 D_k} \, \partial_x\left[ \tilde j_{kz}(x,0) \,  \tilde v_x(x,0)\right] 
\end{eqnarray*}
\bea
&&+ \frac{z^3}{3 D_k}\, \partial_x\left[ \tilde j_{kz}(x,0) \, \partial_z  \tilde v_x(x,0)\right] \nonumber\\
&&+ \frac{z^4}{8\eta D_k}\, \partial_x\left[ \tilde j_{kz}(x,0) \, \partial_x  \tilde P(x,0)\right] + O(q_x^2z)  ,
\label{j-tilde}
\eea
the effective concentration field of species $k$
\be
\tilde c_k(x,z) = \tilde c_k(x,0) - \frac{z}{D_k} \, \tilde j_{kz}(x,0) + O(q_xz^2)  ,
\label{c-tilde}
\ee
the effective pressure
\be
\tilde P(x,z) = \tilde P(x,0) + O(q_xz) ,
\label{P-tilde}
\ee
and the effective velocity field
\be
\tilde v_x(x,z) = \tilde v_x(x,0) + z \, \partial_z \tilde v_x(x,0) + \frac{z^2}{2\eta}\, \partial_x \tilde P(x,0) + O(q_x^2z^2) ,
\label{vx-tilde}
\ee
\bea
&& \tilde v_z(x,z) = \tilde v_z(x,0) -z \, \partial_x \tilde v_x(x,0) - \frac{z^2}{2}\, \partial_x\partial_z \tilde v_x(x,0)\nonumber\\
&&\qquad\qquad\qquad - \frac{z^3}{6\eta} \,
\partial_x^2 \tilde P(x,0) + O(q_x^2z^2) .
\label{vz-tilde}
\eea

According to the matching conditions~(\ref{matching_eqs}), the effective fields must coincide with the actual ones if $z\geq \delta$ where the interaction potentials vanishes.  In particular, this is the case at $z=\delta$ where Eq.~(\ref{A1}) is equal to Eq.~(\ref{j-tilde}), (\ref{A2})~to~(\ref{c-tilde}), (\ref{A3})~to~(\ref{P-tilde}), (\ref{A4})~to~(\ref{vx-tilde}), and (\ref{A5})~to~(\ref{vz-tilde}).  Consequently, we obtain the following relations between the actual and the effective boundary values:
\bea
&&\tilde j_{kz}(x,0) = \nu_k \, w(x) \nonumber\\
&&\qquad\qquad\quad- \partial_x\biggl\{\Big[ K_k^{(0)} \, {\rm e}^{\beta u_k(0)} c_k(x,0)
\nonumber\\
&& -\frac{\nu_k}{D_k}\big(K_k^{(1)}-R_k^{(1)}+S_k^{(0)}\big) w(x) \Big] v_x(x,0)\biggr\}  \nonumber\\
&&\qquad\qquad\quad - \partial_x\biggl\{\Big[ K_k^{(1)} \, {\rm e}^{\beta u_k(0)} c_k(x,0) \nonumber\\
&&-\frac{\nu_k}{D_k}\big(K_k^{(2)}-\frac{1}{2}\, R_k^{(2)}+S_k^{(1)}\big) w(x) \Big] \partial_z v_x(x,0)\biggr\}  \nonumber\\
&&\qquad\qquad\quad - \partial_x\biggl\{\Big[ K_k^{(2)} \, {\rm e}^{\beta u_k(0)} c_k(x,0) \nonumber\\
&&-\frac{\nu_k}{D_k}\big(K_k^{(3)}-\frac{1}{3}\, R_k^{(3)}+S_k^{(2)}\big) w(x) \Big] \frac{1}{2\eta}\, \partial_x P(x,0)\biggr\}  \nonumber\\
&&\qquad\qquad\qquad\qquad\qquad + O(q_x^2\delta)  ,
\label{A-tilde-jkz}
\eea
\bea
\label{ck0-actual-eff}
&&\tilde c_k(x,0) =  {\rm e}^{\beta u_k(0)} c_k(x,0) - \frac{\nu_k}{D_k} \, R_k^{(0)} \, w(x) + O(q_x\delta^2) , \nonumber\\
&&\label{tilde_c-c-w}
\eea
\bea
&&\tilde P(x,0) = P(x,0) + k_{\rm B}T \sum_k  c_k(x,0)\Big[{\rm e}^{\beta u_k(0)}-1 \Big]
\nonumber\\
&&\qquad\qquad\quad - k_{\rm B}T\, w(x) \sum_k\frac{\nu_k}{D_k}\, R_k^{(0)} + O(q_x\delta) ,
\eea
\bea
&& \tilde v_x(x,0) = v_x(x,0)\nonumber\\
&&\qquad\qquad - \frac{k_{\rm B}T}{\eta} \sum_k \partial_x\biggl[ K_k^{(1)} \, {\rm e}^{\beta u_k(0)} c_k(x,0)\nonumber\\
&&\qquad\qquad - \frac{\nu_k}{D_k}\Big(K_k^{(2)}-\frac{1}{2}\, R_k^{(2)}+S_k^{(1)}\Big) w(x) \biggr] \nonumber\\
&&\qquad\qquad\qquad\qquad\qquad + O(q_x^2\delta^2)   ,  \label{vx-vx-tilde}
\eea
\bea
&& \partial_z  \tilde v_x(x,0) = \partial_z v_x(x,0) \nonumber\\
&&\qquad\qquad+ \frac{k_{\rm B}T}{\eta} \sum_k  \partial_x\Big[ K_k^{(0)} \, {\rm e}^{\beta u_k(0)} c_k(x,0) \nonumber\\
&&\qquad\qquad - \frac{\nu_k}{D_k}\big(K_k^{(1)}-R_k^{(1)}+S_k^{(0)}\big) w(x)\Big]\nonumber\\
&&\qquad\qquad\qquad\qquad\qquad+ O(q_x^2\delta)  ,  \label{dvx-dvx-tilde}
\eea
\be
 \tilde v_z(x,0) = O(q_x^2\delta^3)\, ,
\ee
expressed in terms of the quantities~(\ref{K-dfn}), (\ref{R-dfn}), and~(\ref{S-dfn}).

Using the actual boundary condition~(\ref{vx-bc}) together with Eqs.~(\ref{vx-vx-tilde}) and~(\ref{dvx-dvx-tilde}) and expressing the actual fields in terms of the effective ones, we get the boundary conditions for the effective fields given by Eq.~(\ref{new-bc}) with the diffusiophoretic constants~(\ref{b_k}) and the effective concentrations~(\ref{C-dfn}).

Furthermore, Eq.~(\ref{Gk-thin}) shows that the excess surface densities are given by
\bea
&& \Gamma_k(x) = K_k^{(0)} \, \tilde c_k(x,0)\nonumber\\
&& - \frac{\nu_k}{D_k} \Big( K_k^{(1)} - R_k^{(1)}+S_k^{(0)} - K_k^{(0)} R_k^{(0)} \Big) \, w(x)+ O(q_x\delta^2) , \nonumber\\
&& \label{Gamma_k}
\eea
which are of order $\delta$ according to Eq.~(\ref{KRSn-estim}).

In addition, carrying out the integral in Eq.~(\ref{jks-thin}) gives the expression
\bea
&&\int_0^{\delta}(c_kv_x-\tilde c_k \tilde v_x) \, dz = \frac{\lambda\, b_k}{k_{\rm B}T} \, \tilde C_k(x) \, \tilde v_x(x,0) \nonumber\\
&&\qquad\qquad + \frac{1}{2\eta} \, \tilde F_k(x)\, \partial_x \tilde P(x,0) + O(q_x\delta^3) ,
\label{jks-eff}
\eea
with the effective concentrations~(\ref{C-dfn}) and the quantities
\bea
&& \tilde F_k(x) \equiv K_k^{(2)}\tilde c_k(x,0)\nonumber\\
&&  -\frac{\nu_k}{D_k} \big(K_k^{(3)}-\frac{1}{3}\, R_k^{(3)}+S_k^{(2)}-K_k^{(2)} R_k^{(0)} \big) w(x) ,\nonumber\\ &&
\eea
which are of order $\delta^3$.

To evaluate the relative magnitude of the second term with respect to the first in Eq.~(\ref{jks-eff}), let us consider the Poiseuille flow of a solution with uniform concentration gradients of solute species between two chemically inactive planes separated by a distance~$H$.  In this case, the velocity field is of the form
\be
\tilde v_x(x,z) = \frac{\partial_x\tilde P}{2\eta} \, (z^2-Hz-Hb) -\sum_k b_k \partial_x \tilde c_k\, .
\ee
If, moreover, the concentration gradients are small enough and $b=0$, the ratio of the second to the first term in Eq.~(\ref{jks-eff}) takes the value $-K_k^{(2)}/(K_k^{(1)}H)$, which is of order $\delta/H$ according to Eq.~(\ref{KRSn-estim}).  In this regard, the second term can be assumed to be negligible in front of the first.  A similar assumption is considered in Ref.~\onlinecite{AP91}.

Now, the surface velocity can be evaluated as
\be
v_x^{\rm s} = \frac{1}{\delta}\int_0^{\delta} (v_x-\tilde v_x) \, dz =\frac{k_{\rm B}T}{2\eta\delta} \sum_k \partial_x\tilde F_k(x) + O(q_x^2\delta^2) \, ,
\label{vxs}
\ee
which is of order $q_x\delta^2$.  Substituting the different quantities into Eq.~(\ref{jks-thin}), Eq.~(\ref{jks}) is found.

The results also allow us to obtain the surface pressure due to the interaction of the solutes with the wall.  Indeed, for an arbitrarily thin boundary layer, the hydrostatic pressure $P$ in Eq.~(\ref{NS-eq-actual}) corresponds to the combination $\tilde P\theta^+ + P^{\rm s}\delta^{\rm s}$ in terms of the effective pressure $\tilde P$ ruled by Eq.~(\ref{NS-eq-eff}) and the surface pressure $P^{\rm s}$.  Accordingly, this latter is given by
\bea
P^{\rm s}(x) &=& \int_0^{\delta}\left[ P(x,z)-\tilde P(x,z)\right] dz \nonumber\\
&=& k_{\rm B}T \sum_k \Gamma_k(x) + O(q_x\delta^2) \, , \label{Ps-thin}
\eea
as it should.\cite{LL80Part1}


\section{The case of stick boundary condition}
\label{AppB}

Here, we consider the limit of the stick boundary condition ($b=0$). Within the framework of interface nonequilibrium thermodynamics, the surface current density and the tangential components of the pressure tensor are in general related to the velocity slip and the tangential gradients of the surface chemical potentials according to Eqs.~(\ref{BC2}) and~(\ref{BC3}).  Because of the relation $L_{\rm vv} = T  \lambda$ and Eq.~(\ref{L-L}), the right side of Eq.~(\ref{BC3}) increases proportionally to the sliding friction coefficient $\lambda=\eta/b$ in the limit $b=0$ of the stick boundary condition.  In this limit, the boundary value of the pressure tensor exists under the condition that
\be
{\bf v}_{\rm slip}\simeq - \sum_l \frac{b_l c_l}{k_{\rm B}T}\, \pmb{\nabla}_{\bot}\mu_l^{\rm s} \, .
\ee
For consistency between Eqs.~(\ref{BC2}) and~(\ref{BC3}), the relation $L_{kl}/T= \delta_{kl} D_{k}^{\rm s} \Gamma_k/(k_{\rm B}T) -\lambda b_k c_k b_l c_l/(k_{\rm B}T)^2$ should hold.  In the limit $\lambda\to\infty$, there is thus a tight coupling between the surface current density and the tangential pressure tensor
\be
{\bf j}_k^{\rm s} =-\frac{b_k c_k}{k_{\rm B}T} \, {\bf n}\cdot{\boldsymbol{\mathsf P}}\cdot{\boldsymbol{\mathsf 1}}_{\bot} -D_k^{\rm s} \pmb{\nabla}_{\bot}\Gamma_k\, .
\label{tight}
\ee

This result is confirmed within the thin-layer approximation in the presence of surface reaction.
For the stick boundary condition $b=0$, we have that $v_x(x,0) =0$, so that Eq.~(\ref{jks-thin}) here gives the surface current density
\bea
j_{kx}^{\rm s}(x) &=& \frac{\eta \, b_k}{k_{\rm B}T} \, \tilde C_k(x) \,  \partial_z \tilde v_x(x,0) \times \left[1+O(\delta/H)\right] \nonumber\\
&&  -D_k \partial_x\Gamma_k + O(q_x\delta^3)
\label{jks-b=0}
\eea
in terms of the effective concentrations~(\ref{C-dfn}) and the coefficients~(\ref{sigma_k}) with $b=0$, up to corrections that are here also negligible under the conditions $\delta\ll H$ and $q_x\delta\ll 1$.  In addition, the velocity field satisfies the boundary condition~(\ref{new-bc}), which becomes
\be
\tilde v_x(x,0) = - \sum_k b_k \, \partial_x\tilde C_k(x) + O(q_x^2\delta)
\label{v-bc-fin-b=0}
\ee
with the diffusiophoretic constants $b_k = k_{\rm B}TK_k^{(1)}/\eta$ in the limit $b=0$.
Since the tangential component of the pressure tensor is given by
\be
{\bf n}\cdot{\boldsymbol{\mathsf P}}\cdot{\boldsymbol{\mathsf 1}}_{\bot} = -\eta \, \partial_z \tilde v_x(x,0) \, ,
\ee
we find after neglecting the corrections that
\be
j_{kx}^{\rm s}(x) = -\frac{b_k\tilde C_k(x)}{k_{\rm B}T} \,  {\bf n}\cdot{\boldsymbol{\mathsf P}}\cdot{\boldsymbol{\mathsf 1}}_{\bot} -D_k \partial_x\Gamma_k ,
\ee
which is consistent with the expectation~(\ref{tight}) [given the modification~(\ref{C-dfn}) of the concentration by the surface reaction].

Thus, for stick boundary conditions, the surface current density is proportional to the tangential gradient of velocity, instead of the tangential velocity.


%

\end{document}